\documentstyle[11pt,psfig]{article}

\newcommand{\eref}[1]{(\ref{#1})}
\newcommand{\hg}{\hglue.5em plus.5em minus.2em\relax}
\newcommand{\bibref}[5]{\bibitem{#1}{{\sc #2},\hg}{{\it #3\/}}%
                   \hg {{\bf #4},}\hg{{\rm #5}.}}
\begin{document}

\setcounter{topnumber}{5}	
\setcounter{bottomnumber}{5}	
\setcounter{totalnumber}{5}	
\intextsep 3mm		\textfloatsep 3mm	\floatsep 3mm
\renewcommand{\thetable}{\Roman{table}}

\begin{center}

{\LARGE BIFURCATION IN ROTATIONAL \\[3mm]
SPECTRA OF NONLINEAR
   AB$_2$ MOLECULES}

\vspace{3cm}

\begin{large}
Igor N. Kozin\footnote{E-mail: nik@appl.sci-nnov.ru} \\[2mm]
Institute of Applied Physics, \\
 Uljanov Street 46, \\
 603600 Nizhnii Novgorod, Russia \\[5mm]
and \\[5mm]
I.M.\ Pavlichenkov\footnote{E-mail: pavi@kiae.su} \\[2mm]
Russian National Center "Kurchatov Institute",  \\
123182 Moscow, Russia
\end{large}

\end{center}

\pagestyle{plain}

\newpage
\begin{abstract}

A classical microscopic theory of rovibrational motion at high angular
momenta in symmetrical non-linear molecules AB$_2$ is derived within
the framework of small oscillations near the stationary states of a
rotating molecule. The full-dimensional analysis including stretching
vibrations has confirmed the existence of the bifurcation predicted
previously by means of the rigid-bender model [see B.I.\ Zhilinskii
and I.M.\ Pavlichenkov, {\it Opt.\ Spectrosk.\ (USSR)}
{\bf 64}, 413--414 (1988)].
The formation of fourfold energy clusters resulting from the bifurcation
has been experimentally verified for H$_2$Se and
it has been demonstrated in fully-dimensional quantum mechanical
calculations carried out with the MORBID computer program.

We show in the present work that apart from the level
clustering,  the bifurcation produces physically important effects
including molecular symmetry-breaking and a transition from the normal
mode to the local mode limit for
the  stretching vibrations due to rovibrational interaction.
The application of the present theory with realistic molecular potentials
to the H$_2$Te, H$_2$Se and H$_2$S hydrides
results in predictions of the bifurcation points
very close to those calculated previously.
However for the lighter H$_2$O molecule we find that the
bifurcation occurs at higher values of the total angular momentum
than obtained in previous estimations.
The present work shows it to be very unlikely that
the bifurcation in H$_2$O will lead to
clustering of energy levels.
This result is in agreement with recent variational calculations.

\end{abstract}

\newpage
\section{INTRODUCTION}

Energy level clustering in molecules
is very interesting phenomenon, which has been studied
for about twenty years \cite{Dorney,HP,Harter,Sadov}.
However the main interest was  concentrated on highly symmetrical
molecules like tetrahydrides.
The possibility of fourfold cluster formation
in the upper part of rotational multiplets of symmetrical
non-linear triatomic molecules
has been predicted in the theoretical work by one of the authors and
B.I.\ Zhilinskii \cite{ZP}.
In that work the classical rovibrational dynamics of an
asymmetric top molecule
was investigated in the rigid-bender model \cite{houg}.
It has been shown that with the increasing of the angular momentum
quantum number $J$ the local precession around the axis with the
smallest moment of inertia changes to delocalized precession around
two equivalent axes at certain critical value $J_c$.
This phenomenon can be refered to as a bifurcation.
It manifests itself in the softening of the precessional mode, i.e.,
in the decreasing of the spacing between the upper levels of the
rotational multiplets when $J$ passes through the critical value $J_c$.
The noticeable softening of the precessional modes in the ground and
$\nu_2$ vibrational states of H$_2$O and H$_2$S was interpreted as
a possible bifurcation in this molecules \cite{ZP,Pav}.
More precise numerical calculations based on the quantum Hamiltonian
of the rigid-bender model were carried out soon \cite{Mak,Pyka}.
They demonstrated the formation of fourfold energy clusters for $J$
greater that $J_c$.
The microscopic theory of bifurcations developed in Ref.\ \cite{ZP}
allowed the estimation of the  critical angular momentum $J_c$ which
determines the region of  cluster formation.
It was found in Refs.\ \cite{Pav,pav1} that $J_c$ decreases
with increasing mass of the central atom X in the H$_2$X molecule.
This fact has been used in the experimental
study of ``level clustering'' in the rotational spectrum of H$_2$Se
\cite{SUBMM,FIR}.
It was established that the energy levels in the vibrational ground
state of H$_2$Se form groups of four quasi-degenerate levels at the
top of the $J$-multiplets when the value of the total angular momentum $J$
exceeds the critical value $J_c$.
This phenomenon is very unusual for the rotational spectrum of a
rigid asymmetric top.

Soon afterwards the MORBID program has been successfully applied to the
cluster analyses of the spectra of H$_2$Se \cite{JeKo,KoJe},
H$_2$S \cite{H2S} and H$_2$Te \cite{H2Te}.
The MORBID program is a variational approach allowing the calculation of
energy levels of the total rovibrational Hamiltonian of a molecule in
an isolated electronic state \cite{MORBID}.
MORBID calculations based on fitted or {\em ab initio} molecular potentials
showed  good agreement with known experimental data
and predicted the formation of fourfold clusters
in highly excited rotational states of H$_2$Se, H$_2$S and H$_2$Te.
The MORBID calculations were also in good agreement with
pure classical, semiclassical and model quantum mechanical estimations of the
critical angular momentum $J_c$ for these three molecules.
However preliminary calculations \cite{H2Oclu} of the rotational levels
up to $J$ = 42 in the ground vibrational state of water molecule using
the MORBID program with optimized potentials \cite{H2Opot} did not
show evident fourfold clusters. Recently Polyansky {\it et al.} have
reported a new optimized water potential  \cite{Oleg}. Their
calculations up to $J$ = 35 for the ground vibrational state reproduce
very well the available experimental energy levels but the authors
concluded that  the ``rotational energy
level structure in water is at least of a very different nature than
the fourfold cluster structures observed for H$_2$Se and calculated
for H$_2$S, H$_2$Se and H$_2$Te'' \cite{Oleg}.
These two  facts are in disaccord with all estimations of
$J_c$ for the water molecule: $J_c$ = 27$-$28 in Ref.\ \cite{ZP},
$J_c$ = 26 in Ref.\ \cite{Mak}
and with the calculations in Ref.\ \cite{Coudert}, which predict
quasi-degeneracy of  the levels $J_{J,0}$, $J_{J,1}$, $J_{J-1,1}$, and
$J_{J-1,2}$ for $J>$ 32.

This extensive calculations show that the role of stretching
vibrations in the bifurcation phenomenon is not clear enough.
The rigid-bender model
used in the early microscopic theory \cite{ZP} assumes the bond lengths
to be fixed at their equilibrium values.
This theory cannot also answer the question if a bifurcation exists in
the rotational bands of excited vibrational states.
This deficiency is remedied in the present paper,
which uses the fully-dimensional description of rovibrational motion
in a triatomic molecule, the stretching vibrations being included.
The analysis of the classical
equations of motion is based on the study of small harmonic oscillations
near stationary states of the system. Four-dimensional oscillations can be
separated into {\em slow} precessional and {\em fast} vibrational
motion. The approach of the present work differs from the standard
adiabatic approximation by Born and Oppenheimer in that it considers
the rovibrational motion to take place near the
stationary states of a fast uniformly rotating molecule instead
of having them take place near the equilibrium state of a
non-rotating molecule. The difference has very important consequences.
First, the precessional motion of an AB$_2$ molecule around the axis
with minimal moment of inertia becomes unstable at the bifurcation
point $J_c$. For higher angular momenta, the fast uniform rotation of
a molecule takes place around two axes situated in the molecular
plane between the axes of maximal and intermediate moments of inertia.
Owing to molecular symmetry, these two axes are equivalent.
The delocalized precession around them may result in fourfold level
clustering. This effect is the spectroscopic manifestation of the
bifurcation. Another consequence is the asymmetrical
deformation of the AB$_2$ molecule by centrifugal forces, whose
anisotropic action on the B-atoms causes one A---B bond length
to be longer than the other. This symmetry breaking of the molecular
configuration in its turn changes the  vibrational dynamics in such
a way that the normal stretching modes $\nu_1$ and $\nu_3$ transform
to the local vibrations either of two A---B bonds.
It is important to note that
the transition from normal mode to local mode vibrations is caused
in our case by
rovibrational interaction and not by the anharmonicity of the vibrations
as in customary local mode theory
(see, for example, the review article \cite{CH}).
The developed theory allows to understand the relationship of local
mode vibrations to a cluster formation in rotational spectra obtained
by a model consideration in Ref.\ \cite{KKL}.

The classical
 method developed here has been applied to the hydrides H$_2$O, H$_2$S,
H$_2$Se, and H$_2$Te to demonstrate that it produces consistency
with purely quantum mechanical calculations
for quantities like the critical angular momentum,
the bond length distortions and the bending angle of the rotating molecule.
The results obtained for the H$_2$Te, H$_2$Se, H$_2$S molecules
are very close to those obtained previously, but for  the lighter H$_2$O
molecule the bifurcation is found  to occur at $J$ values significantly
higher than those obtained in previous estimations.
The most probable explanation of this is the neglect of  the stretching
vibrations in the earlier models.
Furthermore, our analysis shows also that the stretching vibrations
result in drastic reduction of the potential barrier, which separates
the two regions of a delocalized precession. Thus it is very unlikely
that the bifurcation in  the vibrational ground state of H$_2$O can
result in an observable fourfold cluster structure.
The clustering of levels cannot therefore be
the ultimate criteria of the bifurcation.


\section{CLASSICAL EQUATIONS OF ROVIBRATIONAL MOTION FOR
AN AB$_2$ MOLECULE
 \label{sec:haml} }

We begin with the classical rovibrational Hamiltonian which can be
derived following the method of Wilson, Decius, and Cross \cite{VDC}.
Our treatment differs from theirs in that
we do not use normal coordinates to have the possibility of
considering the large displacements from the equilibrium configuration.
With  the standard choice of Euler angles \cite{VDC}
as rotational coordinates, the classical kinetic rovibrational energy
in a rotating coordinate  system
with the origin in the center of mass of molecule has the form
\begin{equation}
 T = \frac{1}{2} \sum_{ij} I_{ij} \omega_i \omega_j + \sum_i \omega_i g_i
 + \frac{1}{2} \sum_{\alpha i} m_{\alpha} v_{\alpha i}^2,
\label{T}
\end{equation}
where the indices $i$ and $j$ run through the axes $x$, $y$ and $z$ of
this coordinate system,
${\bf r}_{\alpha}(x_{\alpha}, y_{\alpha}, z_{\alpha})$ is the
position vector of nucleus $\alpha$ with mass $m_{\alpha}$,
${\bf v}_{\alpha}$ is its velocity, and ${\bf g} = \sum_{\alpha}
m_{\alpha} ({\bf r}_{\alpha} \times{\bf v}_{\alpha})$.
$I_{ij}$ is a matrix element of the inertia tensor, and
$\omega_i$ are the projections of an angular velocity on the axes
$x$, $y$ and $z$:
\begin{eqnarray}
\omega_x & = & \dot{\theta} \: \sin\chi - \dot{\varphi} \:
 \sin\theta \cos\chi,   \nonumber \\
\omega_y & = & \dot{\theta} \: \cos\chi + \dot{\varphi} \:
 \sin\theta \sin\chi,  \nonumber \\
\omega_z & = & \dot{\varphi} \: \cos\theta + \dot{\chi}.
\end{eqnarray}
We can further introduce the total angular momentum vector  ${\bf J}$
with projections
\begin{eqnarray}
\label{J_i}
J_x & = & p_{\theta} \: \sin\chi - p_{\varphi} \: \csc\theta \cos\chi +
   p_{\chi} \: \cot\theta \cos\chi,   \nonumber \\
J_y & = & p_{\theta} \: \cos\chi + p_{\varphi} \: \csc\theta \sin\chi -
   p_{\chi} \: \cot\theta \sin\chi,   \nonumber \\
J_z & = & p_{\chi},
\end{eqnarray}
which is related to the angular velocity projections as follows \cite{VDC}
\begin{equation}
J_i - g_i = \sum_j I_{ij} \omega_j.
\label{J-g}
\end{equation}
The inversion of Eq.\ \eref{J-g} leads to
\begin{equation}
 \omega_j = \sum_i \mu_{ij} (J_i - g_i),
\label{omega}
\end{equation}
 where $\{\mu_{ij}\}$ is the matrix inverse to $\{I_{ij}\}$.
As a result the kinetic energy takes the form
\begin{equation}
 T = \frac{1}{2} \sum_{ij} (J_i J_j - g_i g_j) \mu_{ij}
 + \frac{1}{2} \sum_{\alpha i} m_{\alpha} v_{\alpha i}^2.
\label{T2}
\end{equation}

We consider a triatomic molecule B$_1$A$_2$B$_3$ with nuclear
masses $m_1$ $=$ $m_3$ $=$ $m$ and  $m_2$ $=$ $M$.
To define               the rotating
(molecule fixed) axis system  we can
use one of the methods suggested by Sutcliffe and Tennyson \cite{ST}.
Let the molecule be placed in the $(xz)$ plane, i.e., $y_i = 0$. Then we
define the $x$-axis to be parallel to the bisector of the bond
angle $\alpha = \widehat{\rm B_1A_2B_3}$.
The axis directions are chosen so that $z$-axis points from B$_3$ to
B$_1$, $x$-axis points from the centre of the mass to A$_2$ and
the coordinate system is right-handed.
It is natural to introduce three internal coordinates:
the A$_2$--B$_1$ distance $r_1 \equiv q_1$,
the angle $\alpha \equiv q_2$ and
the A$_2$--B$_3$ distance $r_3 \equiv q_3$.
These internal coordinates are connected with the cartesian
coordinates $xyz$ as follows
\begin{eqnarray}
x_1 & = & \frac{m r_3 - (M+m)r_1}{M+2m}\cos\frac{\alpha}{2}, \nonumber \\
x_2 & = & \frac{m ( r_1 + r_3)}{M+2m}\cos\frac{\alpha}{2}, \nonumber \\
x_3 & = & \frac{m r_1 - r_3 (M+m)}{M+2m}\sin\frac{\alpha}{2}, \nonumber \\
z_1 & = & \frac{(M+m) r_1 + r_3 m}{M+2m}\sin\frac{\alpha}{2}, \nonumber \\
z_2 & = &-\frac{m( r_1 - r_3 )}{M+2m}\sin\frac{\alpha}{2}, \nonumber \\
z_3 & = &-\frac{(M+m) r_3 + r_1 m}{M+2m}\sin\frac{\alpha}{2}.
\label{xz}
\end{eqnarray}
With a molecule-fixed axis  system defined in this way,
the non-zero components of the inertia-tensor can be written as
\begin{eqnarray}
\label{I_ij}
I_{xz} & = & \frac{m(M+m)}{2(M+2m)}(r_1^2 - r_3^2)\sin\alpha, \nonumber\\
I_{xx} & = & \frac{m}{M+2m} [M(r_1^2 + r_3^2) + m (r_1+r_3)^2]
\sin^2\frac{\alpha}{2}, \nonumber \\
I_{zz} & = & \frac{m}{M+2m} [M(r_1^2 + r_3^2) + m (r_1-r_3)^2]
\cos^2\frac{\alpha}{2}, \nonumber \\
I_{yy} & = & \frac{m(M+m)}{M+2m} (r_1^2 + r_3^2) -
 \frac{2m^2}{M+2m} r_1 r_3 \cos\alpha,
\end{eqnarray}
and the elements of the inverse matrix are
\begin{eqnarray}
 \mu_{xy} & = & \mu_{yz} = 0, \nonumber \\
 \mu_{yy} & = & 1 / I_{yy},   \nonumber \\
 \mu_{xx} & = & I_{zz} / (I_{zz} I_{xx} - I_{xz}^2), \nonumber \\
 \mu_{xz} & = & I_{xz} / (I_{xz}^2 - I_{zz} I_{xx}), \nonumber \\
 \mu_{zz} & = & I_{xx} / (I_{zz} I_{xx} - I_{xz}^2).
\label{mu}
\end{eqnarray}

With the introduced internal coordinates the kinetic energy
\eref{T2} can be transformed to the form
\begin{equation}
 T = \frac{1}{2} \sum_{ij} J_i J_j \mu_{ij}
   + \frac{1}{2} \sum_{\nu\nu'} c_{\nu\nu'} \dot{q}_\nu \dot{q}_{\nu'},
\label{T3}
\end{equation}
where $i$ and $j$ again assume the values $x$, $y$, $z$, and
 $\nu$ and $\nu'$ run through 1, 2, 3.
The $c_{\nu\nu'}$ coefficients have the form
$c_{\nu\nu'}$ = $a_{\nu\nu'} - \sum_{ij} G_{i\nu} G_{j\nu'} \mu_{ij}$,
but for a planar molecule they can be simplified to
$c_{\nu\nu'}$ = $a_{\nu\nu'} - G_{y\nu} G_{y\nu'} \mu_{yy}$.
The matrix elements $a_{\nu\nu'}$ depend on internal
coordinates as follows
\begin{eqnarray}
a_{11} & = & a_{33} = \frac{m(M+m)}{M+2m}, \nonumber \\
a_{22} & = & \frac{m}{4(M+2m)} [ (M+m)(r_1^2+r_3^2) +
  2m r_1 r_3 \cos\alpha ], \nonumber \\
a_{12} & = & a_{21} = \frac{m^2}{2(M+2m)} r_3 \sin\alpha, \nonumber \\
a_{23} & = & a_{32} = \frac{m^2}{2(M+2m)} r_1 \sin\alpha, \nonumber \\
a_{13} & = & a_{31} = - \frac{m^2}{M+2m} \cos\alpha,
\end{eqnarray}
and three $G_{y\nu}$ are written as
\begin{eqnarray}
G_{y1} & = & - \frac{m^2}{2m+M}r_3 \sin\alpha,     \nonumber \\
G_{y2} & = &   \frac{m(m+M)}{2(2m+M)} (r_1^2 - r_3^2 ),  \nonumber \\
G_{y3} & = &   \frac{m^2}{2m+M}r_1 \sin\alpha.
\end{eqnarray}

By introducing conjugate momenta  $p_\nu$ = $\partial T/ \partial
\dot{q}_\nu$ we can write the kinetic energy in the form
\begin{equation}
 T = \frac{1}{2} \sum_{ij} J_i J_j \mu_{ij}
    + \frac{1}{2}  \sum_{\nu\nu'} b_{\nu\nu'} p_\nu p_{\nu'},
\label{T4}
\end{equation}
where $\{b_{\nu\nu'}\}$ = $\{c_{\nu\nu'}\}^{-1}$
assuming Det$\{c_{\nu\nu'}\} \neq 0$.
Finally the rovibrational Hamiltonian of the symmetrical
AB$_2$ molecule is expressed by
\begin{equation}
H = \frac{1}{2} \sum_{ij} J_i J_j \mu_{ij} +
    \frac{1}{2} \sum_{\nu\nu'} b_{\nu\nu'} p_\nu p_{\nu'} +
    V(r_1, \alpha, r_3),
\label{H}
\end{equation}
where $V$ is the nuclear potential, which is symmetrical with respect to
permutations of identical nuclei.

The obtained Hamiltonian allows to write Hamilton's equations of motion
\begin{eqnarray} \label{q_l}
\dot{q}_\nu & = & \sum_{\nu'} b_{\nu'\nu} p_{\nu'}, \\   \label{p_l}
\dot{p}_\nu & = & - \frac{1}{2} \sum_{ij} \frac{\partial \mu_{ij}}
  {\partial q_\nu} J_i J_j - \frac{1}{2} \sum_{\nu'\nu''}
  \frac{\partial b_{\nu'\nu''}}{\partial q_\nu} p_{\nu'} p_{\nu''} -
  \frac{\partial V}{\partial q_\nu}, \\             \label{J_k}
\dot{J_i} & = & \sum_{jkl} e_{ijk} J_j \mu_{kl} J_{l}.
\end{eqnarray}
The equation \eref{J_k} is derived by using the Poisson
bracket $\{J_i, J_j\} = e_{ijk}J_{k}$, where $e$ is asymmetrical
tensor and repeated indexes are summed.
Equation \eref{J_k} is equivalent to the three equations
\begin{eqnarray}
\label{J_x}
\dot{J_x} & = & [ \mu_{xz} J_x + ( \mu_{zz} - \mu_{yy} ) J_z ] J_y,  \\
\label{J_y}
\dot{J_y} & = & ( \mu_{xx} - \mu_{zz} ) J_x J_z +
  \mu_{xz} (J_z^2 - J_x^2 ), \\
\label{J_z}
\dot{J_z} & = & [ -\mu_{xz} J_z + ( \mu_{yy} - \mu_{xx} ) J_x ] J_y.
\end{eqnarray}
They are not independent since
${\bf J}^2$ is an  integral of motion.


\section{PRECESSIONAL AND VIBRATIONAL MOTIONS OF
AN ROTATING AB$_2$ MOLECULE
\label{sec:disc} }

Our analysis of the equations of motion is based on the study of
rotation-vibrational motion near stationary states. In a stationary state,
the time derivatives given in the left hand sides of
Eqs.\ (\ref{q_l}--\ref{J_k}) are equal to zero.
Hence, in a stationary state Eq.\ \eref{q_l} takes the form
\begin{equation}
 \sum_{\nu'} b_{\nu'\nu} p_{\nu'} = 0,
\end{equation}
and it is easy to see that $p_\nu$ = 0, since the determinant of the
$\{b_{\nu\nu'}\}$ matrix is non-zero as mentioned above. Now the right
part of Eq.\ \eref{p_l} can be written in the form
\begin{equation}
\label{p_l:1}
\frac{\partial V_{\rm eff}}{\partial q_\nu} = 0,
\end{equation}
where the effective potential $ V_{\rm eff}$ is given by
\begin{equation}
\label{p_l:2}
V_{\rm eff} = \frac{1}{2} \sum_{ij} \mu_{ij} J_i J_j + V.
\end{equation}
The three Eqs.\ \eref{p_l:1} determine the
configuration (i.e.,  the values of $r_{s1}$, $\alpha_s$, and $r_{s3}$)
of a rotating molecule in a stationary state.
The effective potential  $ V_{\rm eff}$ of Eq.\ \eref{p_l:2}
includes the centrifugal energy, which plays a central role in the
dynamics of a rotating molecule. In particular, the configuration of a
rotating molecule in a stationary state
is different from that of non-rotating one. We will
distinguish the corresponding values by indexes $s$ (for the stationary
state) and $e$ (for the non-rotating molecule in its equilibrium
configuration), respectively.
Although the function $\mu_{xz}$ is antisymmetrical under the permutation
of the coordinates $r_1$ and $r_3$ (see Eq.\ \eref{I_ij}), the effective
potential (Eq.\ \eref{p_l:2}), the total Hamiltonian (Eq.\ \eref{H}), and
the corresponding equations of motion are invariant under the operations
of the C$_{2v}$(M) permutation-inversion group of
the AB$_2$ molecule \cite{Bunker} (see also Ref.\ \cite{JB}).
This group includes identity operator E,
the P(13) permutation operator of identical nuclei, the operator of
inversion E$^*$ and the product P(13)E$^*$ (see Table \ref{t:sym}).

Equations (\ref{J_x}--\ref{J_z}) together with the conservation law for
${\bf J}^2$ define the axis of uniform rotation of  the molecule, i.e.,
the total angular momentum vector ${\bf J}_s$ in the molecular
frame. Three projections of this vector are determined by four algebraic
equations. Therefore, in a stationary state ${\bf J}_s$ cannot have an
arbitrary direction. There are two types of stationary states.
In three {\em axial} stationary states $S_i$: $J_{si} = \pm J$,
$i= x, y, z$, the molecule rotates uniformly around the $i$-axis. The
molecular equilibrium configuration in the state $S_i$ is defined by the
three equations
\begin{equation}  \label{stp}
 \frac{1}{2} \frac{\partial \mu_{ii}}{\partial q_\nu} J^2 +
\frac{\partial V}{\partial q_\nu} = 0
\end{equation}
with $\nu$ = 1, 2, 3.
The molecule has a symmetrical configuration with $r_{s1}$ $=$
$r_{s3}$ $=$ $r_s$ in an axial stationary state S$_i$.
The energy of this state is
\begin{equation}  \label{estp}
E_i = \frac{1}{2} \mu_{ii}(r_s,\alpha_s,r_s) J^2 + V(r_s,\alpha_s,r_s).
\end{equation}

There is only one {\em plane} stationary state $S_{xz}$.
In this state, the molecule rotates uniformly
around one    of the two equivalent axes situated in the
$(xz)$ plane symmetrically relative to the $x$-axis. They form angles
$\beta_s$ and $(\pi - \beta_s)$ with the $z$-axis.
The configuration of the molecule and the value of
$\beta_s$ are defined by the equations
\begin{eqnarray} \label{4stp_b}
& &
  \frac{1}{2} \left(
 \frac{\partial \mu_{xx}}{\partial q_\nu} \sin^2 \beta_s +
 \frac{\partial \mu_{xz}}{\partial q_\nu} \sin 2 \beta_s +
 \frac{\partial \mu_{zz}}{\partial q_\nu} \cos^2 \beta_s \right)J^2 +
 \frac{\partial V}{\partial q_\nu} = 0, \qquad \nu = 1,2,3  \\
\label{4stp_c} & &
  \frac{1}{2} ( \mu_{xx} - \mu_{zz} ) \sin 2 \beta_s +
    \mu_{xz} \cos 2 \beta_s = 0.
\end{eqnarray}
It is interesting that in the state $S_{xz}$, the molecule
has an asymmetric  configuration with $r_{s1} \not= r_{s3}$.
Of the two bond lengths, the one which initially forms the largest angle
with the rotation axis will experience a larger elongation due to
centrifugal forces, and for increasing $J$ this bond length will
tend to become perpendicular to the rotation axis.
When we have obtained one solution of Eqs.\ (\ref{4stp_b}--\ref{4stp_c}),
the remaining three solutions can be easily obtained by using
symmetry operations of the C$_{2v}$(M) group from Table \ref{t:sym}.
The energy of the molecule in the $S_{xz}$ state is equal to
\begin{equation}
\label{exz}
E_{xz} = \frac{1}{2}J^2(\mu_{xx} \sin^2 \beta_s + \mu_{xz} \sin 2 \beta_s +
  \mu_{zz} \cos^2 \beta_s) + V(r_{s1},\alpha_s,r_{s3}).
\end{equation}

The analysis of stationary states shows that an axial state has higher
symmetry in the C$_{2v}$(M) group than the plane state.
This means that transitions from axial to plane state or vice versa
are accompanied by a C$_{2v}$-type bifurcation \cite{pav1}.
Such a bifurcation has been considered for the first time in rotational
spectra of symmetrical triatomic molecules in Ref.\ \cite{ZP}.
However, with the rigid-bender model with fixed bond lengths used in that
work, it was not possible to show that the bifurcation results not
only in the  splitting of the stationary axis and consequently in
level clustering but also in molecular symmetry-breaking.
Due to this latter effect, the vibrational dynamics of the rotating
molecule changes as we shall see below.
Table \ref{tabl1} shows the centrifugal distortions
$\Delta r_\nu$ = $ r_{s\nu}$ $-$ $r_e $ of the bond lengths
in the stationary state $S_{xz}$.
The values were obtained numerically for  the molecules H$_2$Se and H$_2$S
by using the potentials from Refs.\ \cite{JeKo} and \cite{H2S},
respectively.
For comparison, the $\Delta r_\nu$ values obtained for the quantum
cluster states in the upper part of the $J$-multiplets belonging to the
ground vibrational state of these molecules are given in the same
table. It is seen from Table \ref{tabl1} that classical and quantum
estimates are in a good agreement.

Since the derivative $\frac{\partial\mu_{zz}}{\partial\alpha}$ is
positive in the stationary state $S_z$, the bending angle $\alpha_s$
decreases with increasing $J$ according to Eq.\ \eref{stp}.
After a bifurcation point, the angle $\alpha_s$ is stabilized
at a constant value in the stationary state $S_{xz}$.
In the rigid-bender model this angle is independent of $J$ for $J>J_c$
\cite{ZP,Pav} (see Fig.\ \ref{f:angle}). For H$_2$Se, the rigid-bender
and stretching-bender models give comparable results, but for H$_2$O
the predictions of the stretching-bender model deviate significantly from
those of the rigid-bender (Fig.\ \ref{f:angle}). In the stretching-bender
model, the bending angle value for H$_2$O continues
to decrease significantly when $J$ is above the critical value.
Hence no stabilization at a constant value takes place.
This effect can be explained as a competition between centrifugal force
causing the hydrogen nuclei to bring together and their mutual repulsion.

In order to follow the change of molecule rotational regimes as $J$
increases, we must investigate the stability of the stationary states.
To do this we consider the linearized set of Eqs.\ (\ref{q_l}--\ref{J_k})
for small displacements of the internal coordinates
$Q_\nu = q_\nu - q_{s\nu}$ and  the angular momentum projections
$J'_i = J_i - J_{si}$ from their stationary values.
Let us begin with the axial stationary state $S_y$. The linearized
equations have the form
\begin{eqnarray} \label{e:liny}
& &
\sum_{\nu'}\left(c_{\nu\nu'} \ddot{Q}_{\nu'} + \frac{\partial^2 V_{\rm eff}}
  {\partial q_\nu  \partial q_{\nu'}} Q_{\nu'}\right) = 0, \nonumber \\
& &
\ddot{J}_x + J^2 (\mu_{xx} - \mu_{yy}) (\mu_{zz} - \mu_{yy}) J_x = 0.
\end{eqnarray}
The coefficients of these equations are assumed to be taken in the
stationary state, i.e., for $q_\nu = q_{s\nu}$ (see Eqs.\ \eref{stp}).
Note also
that $J'_x = J_x$ because of $J_{sx} = 0$ in the considered stationary
state. Equations \eref{e:liny} describe two independent harmonic motions:
vibrational and precessional. The latter motion is characterized by the
precession frequency
\begin{equation}
\label{omega_y}
\Omega_y = J [(\mu_{xx} - \mu_{yy}) (\mu_{zz} - \mu_{yy})]^{1/2},
\end{equation}
which has the same form as for a rigid top \cite{land}. The vibrational motion
in the state $S_y$ does not differ from that of non-rotating molecule except
in that
the frequencies $\omega_1$, $\omega_2$, and $\omega_3$ of three
normal modes
$\nu_1$, $\nu_2$, and $\nu_3$ \cite{VDC} are determined by the effective
potential and the equilibrium configuration of rotating molecule. As a
result they are shifted relative to those of non-rotating molecule
(see below).

The linearized equations for the stationary state $S_z$
\begin{eqnarray} \label{e:linz}
& &
\sum_{\nu'}\left(c_{\nu\nu'} \ddot{Q}_{\nu'} + \frac{\partial^2 V_{\rm eff}}
  {\partial q_\nu  \partial q_{\nu'}} Q_{\nu'}\right) +
  J\frac{\partial \mu_{xz}}{\partial q_\nu} J_x = 0,   \nonumber \\
& &
\ddot{J}_x + J^2 (\mu_{zz} - \mu_{yy}) \left\{ (\mu_{zz} - \mu_{xx}) J_x -
   J\sum_\nu \frac{\partial \mu_{xz}}{\partial q_\nu} Q_\nu \right\} = 0,
\end{eqnarray}
describe coupled vibrational and precessional motions. It is not
difficult to see that the coupling originates in the antisymmetrical
variable $Q_a = (Q_1 - Q_3)/\sqrt{2}$. Thus Eqs.\ \eref{e:linz} split
into two independent parts for the variables $Q_a$, $J_x$ and
$Q_s = (Q_1 + Q_3)/\sqrt{2}$, $Q_2$. The equations for $Q_s$ and $Q_2$ are
independent of  the precessional motion
and describe the $\nu_1$ and
$\nu_2$ vibrational modes modified by molecular rotation. We can
separate the precessional motion
from the antisymmetric vibration by using
the inequality $\omega_3 \gg \Omega_z$ ({\em adiabatic approximation}),
which is correct up to  the critical point $J_c$ (see below). In this
approximation, the frequency of molecular precession around the $z$-axis
is equal to
\begin{equation}
\label{omega_z}
\Omega_z = J\left[(\mu_{zz} - \mu_{yy}) \left\{\mu_{zz} - \mu_{xx} +
 J^2\left(\frac{\partial\mu_{xz}}{\partial r_a}\right)^2\bigg/
 \frac{\partial^2 V_{\rm eff}}{\partial r_a^2}\right\}\right]^{1/2},
\end{equation}
where according to Eqs.\ \eref{I_ij} and \eref{mu}
\begin{equation} \label{e:mzz-mxx}
\mu_{zz}-\mu_{xx} = \frac{2 (M+m)}{mMr_s^2 \sin \alpha_s}
   \left(\frac{m}{M+m} -\cos \alpha_s \right).
\end{equation}
For the hydrides considered, the equilibrium angle $\alpha_e$ is larger
than 90$^\circ$. So the precession frequency of Eq.\
\eref{omega_z} is real and the
stationary state $S_z$ is stable for small values of $J$. As we saw
above, the stationary angle $\alpha_s$ decreases with increasing $J$ due
to a centrifugal distortion. So $\mu_{zz}$ approaches
$\mu_{xx}$ and the precession frequency $\Omega_z$ vanishes when
\begin{equation}
\label{J_c}
\mu_{zz} - \mu_{xx} + J_c^2\left(\frac{\partial\mu_{xz}}{\partial r_a}
\right)^2\bigg/\frac{\partial^2 V_{\rm eff}}{\partial r_a^2}=0.
\end{equation}
Equations \eref{J_c} and        \eref{stp} (for $i=z$) define
the critical angular momentum $J_c$. In the
rigid-bender model, this point is determined by the bending angle
\begin{equation}
\label{alpha_c}
\alpha_c = \arccos \left( \frac{m}{M+m} \right),
\end{equation}
for which $I_{xx} = I_{zz}$. The last term in Eq.\ \eref{J_c} takes into
account the non-rigidity of the bond lengths $r_1$ and $r_3$ and becomes
important when $\mu_{zz} - \mu_{xx}$ is close to zero. Since the
correction term is
positive, Eq.\ \eref{J_c} gives  higher values of $J_c$ than the
rigid-bender model. Clearly, the more rigid the antisymmetrical stretching
mode $\nu_3$, the closer is $J_c$ to the value
obtained in Ref.\ \cite{Pav} in the rigid-bender approximation. Our
numerical calculations of critical angular momentum for the
set of the H$_2$X hydrides with the potentials from Refs.\
\cite{JeKo,H2S,H2Te} give mainly slight corrections to the $J_c$
values predicted previously.
 We obtained $J_c$=9.3 for H$_2$Te, $J_c$=12.5 for H$_2$Se and
$J_c$=18.9 for H$_2$S compared to rigid-bender $J_c$ values of
8.5, 11.4, and 16.9, respectively.
It is seen that for lighter molecules, larger discrepancies are obtained.
An exception is the water molecule
for which the discrepancy of the $J_c$ value is very significant.
Using the potential from Ref.\ \cite{H2Opot} $J_c$=26.5 was obtained with
 the ``frozen''
bond lengths (which is close to all
previous     estimations) and $J_c$=35.2 in
the stretching-bender model.

For $J>J_c$, the stationary state $S_z$
corresponds to an unstable saddle-type point.
It is the top of a separation barrier between two symmetrical
states $S_{xz}$ which appear because of the bifurcation.
The equations of a small amplitude motion near this state have the form
\begin{eqnarray} \label{e:linxz}
& &
\sum_{\nu'}\left(c_{\nu\nu'} \ddot{Q}_{\nu'} + \frac{\partial^2 V_{\rm eff}}
  {\partial q_\nu  \partial q_{\nu'}} Q_{\nu'}\right) +
  \frac{J B_\nu}{\cos\beta_s} J'_x  = 0,   \nonumber \\
& &
\ddot{J}'_x + J^2 (\mu_{zz} - \mu_{yy} + \mu_{xz} \tan{\beta_s})
   \left( \frac{\mu_{zz} - \mu_{xx}}{\cos{2\beta_s}} J'_x -
   J \cos\beta_s \sum_\nu B_\nu Q_\nu \right)  = 0,
\end{eqnarray}
where the ``coupling constant''
\begin{equation}
\label{coupl}
B_\nu = \frac{1}{2} \frac{\partial(\mu_{xx} - \mu_{zz})}{\partial q_\nu}
\sin{2\beta_s} + \frac{\partial\mu_{xz}}{\partial q_\nu} \cos{2\beta_s},
\end{equation}
mixes all four motions: three vibrations and a precession. The last one
can be decoupled from  the vibrations in the adiabatic approximation for
moderate values of $J$.
In this approximation, the frequency of the precession
around the $S_{xz}$ state is
\begin{equation}
\label{omega_xz}
\Omega_{xz} = J \left[(\mu_{zz} - \mu_{yy} + \mu_{xz} \tan \beta_s )
  \left( \frac{\mu_{zz} - \mu_{xx}}{\cos 2\beta_s} + \frac{J^2}{\Delta}
  \sum_{\nu,\nu'}(-1)^{\nu - \nu'} \Delta_{\nu\nu'} B_\nu B_{\nu'} \right)
  \right]^{1/2},
\end{equation}
where the determinant
$\Delta = {\rm Det}\left\{\frac{\partial^2 V_{\rm eff}}
{\partial q_\nu \partial q_{\nu'}}\right\}$ and
$\Delta_{\nu\nu'} = \Delta_{\nu'\nu}$ are the minores of this determinant.
All the values in
Eq.\ \eref{omega_xz} are assumed to be taken in the stationary point
determined by Eqs.\ \eref{4stp_b} and \eref{4stp_c}. It is not difficult
to show that the precession frequency $\Omega_{xz}$ vanishes in the
critical point $J_c$.

The small amplitude motion near the state $S_x$ is described by
Eqs.\ \eref{e:linz} when   index $x$ is        replaced by $z$.
As a result we get
in the adiabatic approximation the following expression for the
precession frequency around the $x$-axis
\begin{equation}
\label{omega_x}
\Omega_x = J\left[(\mu_{xx} - \mu_{yy}) \left\{\mu_{xx} - \mu_{zz} +
 J^2\left(\frac{\partial\mu_{xz}}{\partial r_a}\right)^2\bigg/
 \frac{\partial^2 V_{\rm eff}}{\partial r_a^2}\right\}\right]^{1/2},
\end{equation}
where all the values are taken in the stationary state $S_x$. The last
term in the  braces   is a small correction. Consequently, the
stability of the $S_x$ state is mainly defined by the value
given in Eq.\ \eref{e:mzz-mxx}
which in turn depends on the angle $\alpha_s$.
Since this angle is larger than 90$^\circ$ for  the  molecules considered
and increases with increasing $J$ because the derivative
$\frac{\partial \mu_{xx} }{\partial \alpha}$ is negative,
the precession frequency $\Omega_x$ is imaginary for reasonable
values of $J$.
Thus the state $S_x$ corresponds to an unstable saddle-type point.
The $J$-dependence of its energy is defined by Eq.\ \eref{estp}.

Equations \eref{estp}
and \eref{exz} (when the transition $S_{z}\to S_{xz}$ occurs)
provide a good estimation of the maximal rotation energy in $J$-multiplet.
The difference between our estimations and experimental data or
variational calculations for H$_2$Te, H$_2$Se, H$_2$S and H$_2$O
molecules does not exceed 10\%\ . This is also true for our estimations
of the minimal rotational energy in the $J$-multiplet using Eq.\
\eref{estp} for the state $S_y$. The energy of the unstable stationary
state $S_x$ corresponds to the top of the barrier responsible for the
inversion doublets ($K$-doublets)
in the upper and lower parts of the $J$-multiplets \cite{Harter}.
Another barrier appears because of the bifurcation for
angular momentum at $J>J_c$. It separates two degenerate maxima of
the stationary state $S_{xz}$. The barrier is responsible for the
twofold clustering of the inversion doublets. Its height is equal to the
difference between the energies of  the states $S_z$ and $S_{xz}$.
The higher the barrier is, the more apparent is  the clustering.
In the H$_2$Te, H$_2$Se and  H$_2$S molecules, the height of the barrier
becomes significant as $J$ increases. However,
calculations carried for the H$_2$O molecule with potentials from Ref.\
\cite{H2Opot} show that the height of the energy barrier increases very
slowly with increasing $J$ and its value is only 15~cm$^{-1}$ at $J$=40
(see Fig.\ \ref{f:h2o}). This value is less than
the precession frequency in the $S_{xz}$ state so that
such a barrier cannot result in fourfold clustering of levels.
For higher $J$ values the barrier increases further to its maximal value
150~cm$^{-1}$ (which is about the precession frequency at this point)
and then begins to decrease.
In contrary, the barrier rapidly increases with increasing $J$
for the rigid-bender model
as it is seen from Fig.\ \ref{f:h2o}, where the
dependence of the barrier height on angular momentum $J$ is plotted.
It is obvious that water represents a special
case in the series of H$_2$X hydrides.
Despite the bifurcation in the ground vibrational state
of this molecule, no fourfold level clustering can be observed
in the upper part of $J$-multiplets since the barrier is too small.
This fact gives us a physical explanation to the recent numerical
calculations of water rotational spectrum with highly accurate optimized
potentials \cite{H2Oclu,Oleg}, which failed to show any fourfold clusters.

The C$_{2v}$ bifurcation has to exist also in rotational energy
structure of
excited vibrational states so far as the adiabatic approximation
$\omega \gg \Omega$ is valid and allows to separate the precession
from the vibrational motion.
As to the energy clustering effect, we have seen above that this
problem requires separate and more careful consideration.
Preliminary analysis of experimental data showed the tendency
of levels to cluster in $\nu_2$ vibrational states of the H$_2$O and
H$_2$S molecules \cite{ZP,Pav}.
The cluster phenomenon has been found in variational
quantum calculations for the
excited vibrational states of H$_2$Se \cite{JeKo,KoJe}, H$_2$S \cite{H2S}
and H$_2$Te \cite{H2Te} and very recently
observed experimentally in
the $\nu_1/\nu_3$ state of H$_2$Se \cite{Flaud}.

Consider now in detail the molecular vibrational motion near the
stationary states $S_z$ for $J<J_c$ and $S_{xz}$ for $J>J_c$. Four normal
modes of precession-vibration motion are described by four coupled
Eqs.\ \eref{e:linz} and \eref{e:linxz} correspondingly. All these equations
are not correct near the critical point $J_c$ where the rovibrational
motion is strongly    non-linear. It was shown above that the equations
for the axial state $S_z$ (and those for the state $S_x$) can be separated
into two independent pairs when symmetrized stretching
coordinates are introduced. When $J$= 0 the precessional motion is absent and
the vibrational motion corresponds to the well-known picture of the normal
vibrational modes $\nu_1$, $\nu_2$ and $\nu_3$ \cite{VDC}. This limit
allows us to refer to the standard notations for a rotating molecule as well,
i.e., the frequency $\omega_1$ corresponds to the symmetrical stretching
mode $\nu_1$, the frequency $\omega_2$ corresponds to the bending mode
$\nu_2$, and the frequency $\omega_3$ corresponds to the antisymmetrical
stretching mode $\nu_3$. The difference from a non-rotating molecule
consists in the effective potential for vibrations and in the equilibrium
configuration, which is symmetrical but differs from that of non-rotating
molecule. Besides, the precessional mode with frequency $\Omega_z$
appears and the antisymmetrical stretching vibration mixes with the
precessional motion
because of a rovibrational interaction. However this mixing is very weak.
This can be seen from Table \ref{tabl2} where the normalized
amplitudes of four coordinates  $Q_1$, $Q_2$, $Q_3$, and
$Q_4= J_x/J$
involved in a normal mode oscillations are given for the H$_2$Se molecule
at $J$= 10. It is seen from the table that the mixing of different types
of motions is small and the character of the vibrations is close
to the standard normal mode picture \cite{VDC}.

The character of the oscillations changes drastically after the critical
point $J_c$.
It is impossible to split the four equations \eref{e:linxz} describing
the small amplitude motion near the stationary state $S_{xz}$ into two
independent pairs because the molecular configuration is asymmetric and
the uniform rotation of molecule around the
axis in the $(xz)$ plane mixes the symmetrical and antisymmetrical
stretching vibrations. As a result, the standard picture of normal mode
vibrations, which is correct for the $S_z$ state, breaks down. This
effect is illustrated in Table \ref{tabl3} which shows normalized
amplitudes of oscillations around the asymmetrical configuration of the
state $S_{xz}$ at $J$=16 ($\beta_s$ = 26.08$^\circ$).
It shows also that the
precessional motion is mixed mostly with the bending vibration.
It is worth noting that the vibrations in the mode with the frequency
$\omega_1$ are localized predominantly on the bond $r_1$ and
vibrations in the
mode with the frequency $\omega_3$ are localized on the bond $r_3$.
This feature is inherent in local mode vibrations \cite{CH}.
However usually the local mode motion appears due to the strong
 anharmonicity as the stretching vibrations are increasingly excited.
In our case the normal to local mode transition is a result of
the symmetry breaking, which leads
to the different force   constants for two bonds and favors the local
mode picture in the competition between kinetic and potential coupling
between the local bond oscillators.
Accordingly the local character of the stretching motion strengthens with
increasing $J$. This follows from Table \ref{tabl4} which shows normalized
amplitudes of oscillations around
the state $S_{xz}$  at $J$=40 ($\beta_s$ = 42.26$^\circ$).

The classical calculations of the present work have shown
that in non-linear AB$_2$ molecules,
a normal to local mode transition can be brought about by rotational
excitation. Similar results have been obtained
in quantum numerical calculations with the MORBID
program \cite{KoJe}. The classical analysis of precession-vibrational
motion of the present work gives a physical picture of the phenomenon.
That is, the bifurcation in a ground vibrational state
breaks the molecular symmetry and changes
both rotational and vibrational motions of the non-linear
AB$_2$ molecule resulting in the fourfold clusters and in the normal to
local mode transition.
The physical reasons for these phenomena lie in the molecular
symmetry and in the anisotropic centrifugal force. Therefore we should
expect similar or even more interesting phenomena in the fast
rotating molecules with higher symmetry.


\section{ACKNOWLEDGMENTS}

This work was supported by the Russian Fond for Fundamental
Investigations (through grant N 94-02-05424-a), International
Science Foundation and Russian Government (through grants MIW000,
MIW300 and R8I000, R8I300). The authors are indebted to P.\ Jensen
for critically reading the manuscript and suggesting improvements.
I.N.K.\ is thankful to A.F.\ Krupnov for valuable comments on the manuscript.


\newcommand{\JMS}{J.~\ Mol.\ Spectrosc.}

\clearpage

{\Large FIGURE CAPTIONS}
\vspace{2cm}

\begin{figure}[h]

\caption{ \label{f:angle}
$J$-dependence of the bending angle
in the stationary states with maximal rotational energy ($S_z$ if
$J<J_c$ and $S_{xz}$ if $J>J_c$) for the H$_2$Se and H$_2$O molecules.
The angle is displayed relative to the equilibrium value $\alpha_e$ for
non-rotating molecule. The solid lines correspond to the
stretching-bender model with the critical angular momentum $J_c$=12.5
(H$_2$Se) and 35.2 (H$_2$O);
the dashed lines represent the rigid-bender model
($J_c$=11.4 for the H$_2$Se and 26.5 for the H$_2$O molecules).}

\end{figure}

\begin{figure}[h]

\caption{ \label{f:h2o}
Value of the barrier which separates two equivalent maxima of stationary
state $S_{xz}$ for water molecule calculated in the stretching-bender model
(solid line) and the rigid-bender model (dashed line).}

\end{figure}

\clearpage

{\Large TABLE CAPTIONS}
\vspace{2cm}

\begin{table}[h]

\caption{ \label{t:sym}
Action of C$_{2v}$(M) group operators on the internal coordinates and
the projections of the total angular momentum $\bf J$.}

\end{table}

\begin{table}[h]

\caption{ \label{tabl1}
Comparison of the molecular centrifugal distortions $\Delta r_\nu = r_{s\nu}
- r_{e\nu}$ for the stationary state $S_{xz}$ at $J$=20 calculated in
quantum and classical methods. The classical calculations correspond to
molecular rotation around the stationary axis which is approximately
perpendicular to the bond $r_1$.
Indexes $s$ and $e$ stand for the molecular equilibrium
configurations of rotating and non-rotating molecule correspondingly.}

\end{table}

\begin{table}[h]

\caption{ \label{tabl2}
Normalized amplitudes of coordinate oscillations in the normal modes
of precession-vibrational motion ($Q_4 = J_x/J$) around stationary
state $S_z$ for H$_2$Se molecule at $J$=10.}

\end{table}

\begin{table}[h]

\caption{ \label{tabl3}
Normalized amplitudes of coordinate oscillations in the normal modes
of precession-vibrational motion ($Q_4 = J'_x/J_{sz}$) near the equilibrium
configuration of the H$_2$Se molecule uniformly rotating in
stationary state $S_{xz}$ at $J$=16 around the axis which is approximately
perpendicular to the $r_1$ bond ($\beta_s$= 26.08$^\circ$).}

\end{table}

\begin{table}[h]

\caption{ \label{tabl4}
Normalized amplitudes of coordinate oscillations in the normal modes
of precession-vibrational motion ($Q_4 = J'_x/J_{sz}$) near the equilibrium
configuration of the H$_2$Se molecule uniformly rotating in
stationary state $S_{xz}$ at $J$=40 around the axis which is approximately
perpendicular to the $r_1$ bond ($\beta_s$= 42.26$^\circ$).}

\end{table}

\begin{figure} [p]
Figure 1.

\centerline{\protect\psfig{figure=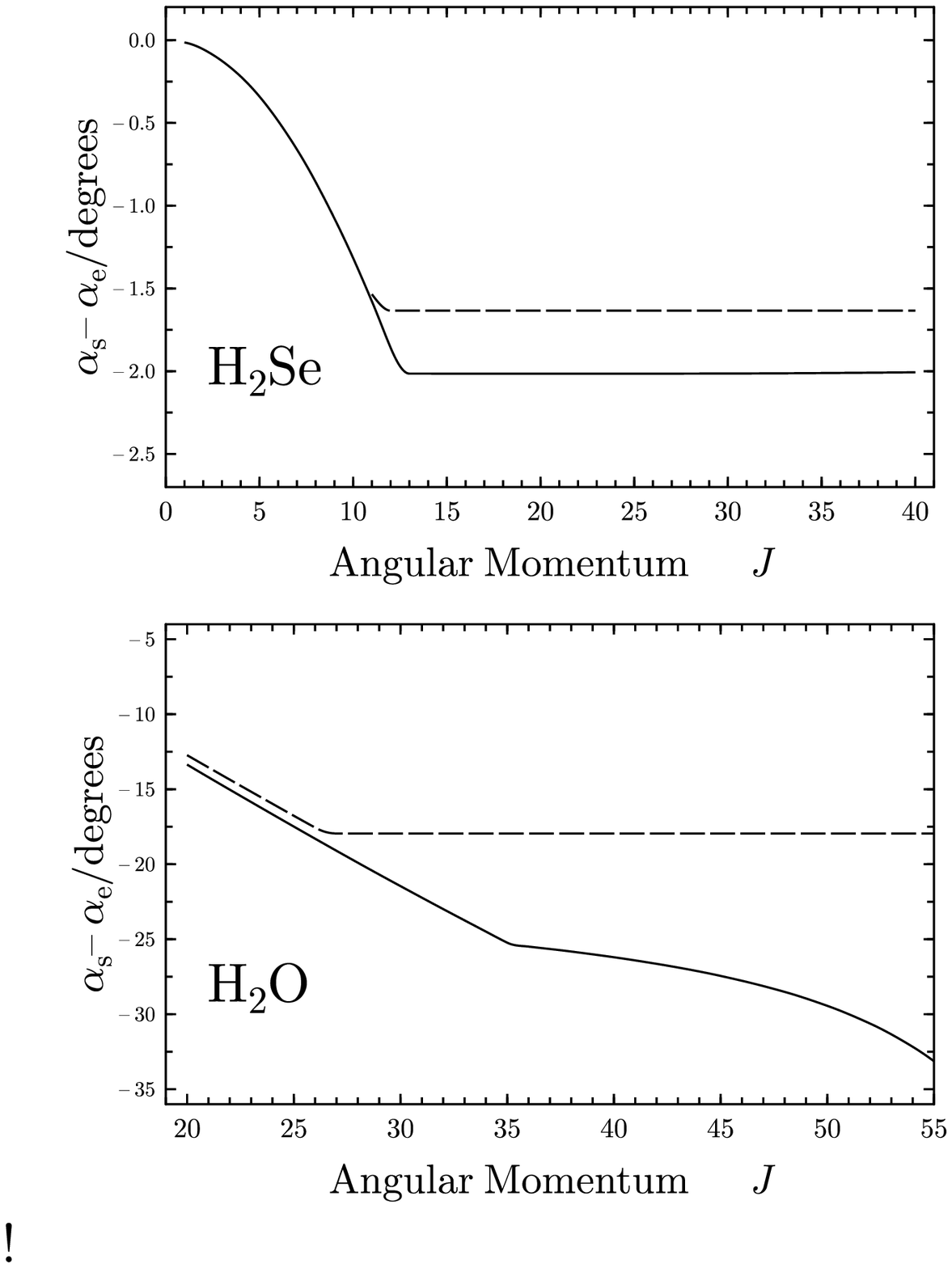,width=6.0in}}
\end{figure}

\newpage
\begin{figure} [p] 
Figure 2.

\centerline{\protect\psfig{figure=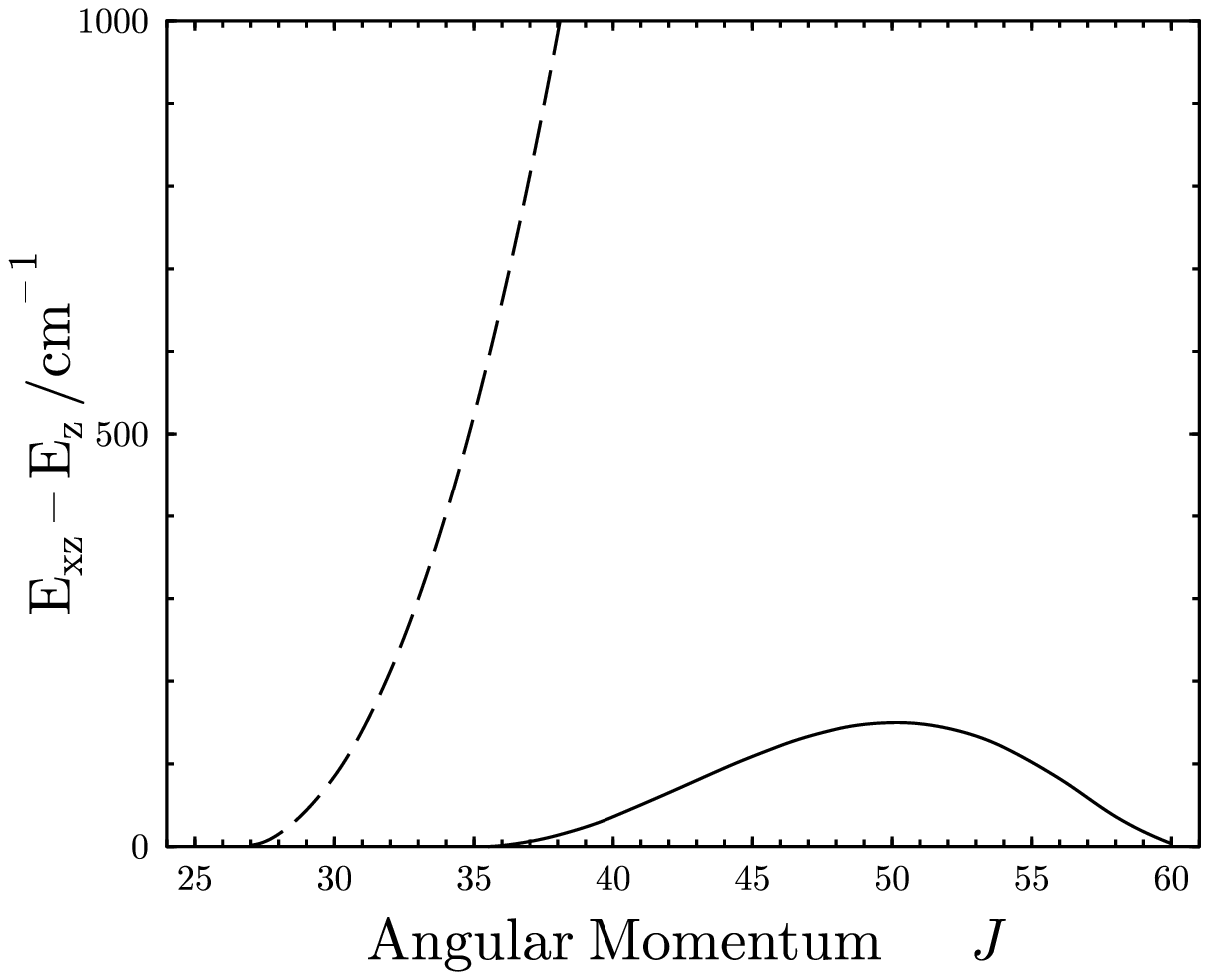,width=5.0in,angle=270}}
\end{figure}

\clearpage
\begin{table} [htb]
Table \ref{t:sym}. \\[1cm]
 \begin{center}
  \begin{tabular}{|c|cccc|} \hline \hline
	 &   E    &  P(13)  &  E$^*$  &  P(13)E$^*$ \\
\hline
 $r_1$   & $r_1$  & $r_3$   & $r_1$   & $r_3$   \\
 $r_3$   & $r_3$  & $r_1$   & $r_3$   & $r_1$   \\
$\alpha$ &$\alpha$& $\alpha$& $\alpha$& $\alpha$  \\
 $J_x$   & $ J_x$ &  $ J_x$ &  $-J_x$ &  $-J_x$  \\
 $J_y$   & $ J_y$ &  $-J_y$ &  $ J_y$ &  $-J_y$  \\
 $J_z$   & $ J_z$ &  $-J_z$ &  $-J_z$ &  $ J_z$   \\
\hline \hline
  \end{tabular}
 \end{center}
\end{table}

\begin{table} [htb]
Table \ref{tabl1}. \\[1cm]
 \begin{center}
  \begin{tabular}{|c|cc|cc|}
\hline \hline
molecule  & \multicolumn{2}{|c}{quantum estimation}
& \multicolumn{2}{|c|}{classical estimation}  \\
	  & $\Delta r_1$/ \AA & $\Delta r_3$/ \AA
& $\Delta r_1$/ \AA & $\Delta r_3$/ \AA \\
\hline
H$_2$Se   &  0.023$^a$ & 0.004$^a$ & 0.025 & 0.002 \\
H$_2$S    &  0.024$^b$ & 0.009$^b$ & 0.022 & 0.010 \\
\hline \hline
  \end{tabular}
 \end{center}
$^a$ --- taken from Ref.\ \cite{KoJe}. \\
$^b$ --- taken from Ref.\ \cite{H2S}.
\end{table}

\newpage

\begin{table} [htb]
Table \ref{tabl2}. \\[1cm]
 \begin{center}
  \begin{tabular}{|rr|r@{.}l|r@{.}l|r@{.}l|r@{.}l|}
\hline \hline
\multicolumn{2}{|c}{ Normal} & \multicolumn{8}{|c|}{Normalized amplitudes} \\
   \cline{3-10}
\multicolumn{2}{|c}{ frequency/cm$^{-1}$}
   & \multicolumn{2}{|c}{\rule{0em}{1.5em} $(r_1-r_{s1})/$\AA}
   & \multicolumn{2}{|c}{$(\alpha-\alpha_s)/$rad.}
   & \multicolumn{2}{|c}{$(r_3-r_{s3})/$\AA}
   & \multicolumn{2}{|c|}{$J_x/J$} \\
\hline
 $\omega_1$=& 2426.4 &   1 &$^\dagger$ &$-$0 & 032 &   1 &   &   0 &  \\
 $\omega_2$=& 1082.3 &$-$0 & 014 & 1 &$^\dagger$ &$-$0 & 014 &   0 &  \\
 $\omega_3$=& 2437.4 &   1 &$^\dagger$ &   0 &     &$-$1 &     &$-$0 & 004\\
 $\Omega_z$=&   15.7 &   0 & 004 &   0 &     &$-$0 & 004 &  1 &$^\dagger$  \\
\hline \hline
  \end{tabular}
 \end{center}
$^\dagger$ --- Normalization condition $Q_n$=1.
\end{table}

\vspace{2cm}

\begin{table} [htb]
Table \ref{tabl3}. \\[1cm]
 \begin{center}
  \begin{tabular}{|rr|r@{.}l|r@{.}l|r@{.}l|r@{.}l|}
\hline \hline
\multicolumn{2}{|c}{ Normal } & \multicolumn{8}{|c|}{Normalized amplitudes} \\
    \cline{3-10}
\multicolumn{2}{|c}{ frequency/cm$^{-1}$}
   & \multicolumn{2}{|c}{\rule{0em}{1.5em} $(r_1-r_{s1})/$\AA}
   & \multicolumn{2}{|c}{$(\alpha-\alpha_s)/$rad.}
   & \multicolumn{2}{|c}{$(r_3-r_{s3})/$\AA}
   & \multicolumn{2}{|c|}{$(J_x-J_{sx})/J_{sz}$} \\
\hline
 $\omega_1$=& 2390.9 &  1 &$^\dagger$ &0 & 018 & 0 & 095 &    0 & 003 \\
 $\omega_2$=& 1110.1 &$-$0 & 015 &1 &$^\dagger$  & $-$0 & 013 &    0 & 020 \\
 $\omega_3$=& 2436.0 &$-$0 & 095 &0 & 012 & 1 &$^\dagger$ & $-$0 & 003 \\
 $\Omega_{xz}$=&  39.5 &   0 & 006 &0 & 081 & $-$0 & 008 & 1 &$^\dagger$  \\
\hline \hline
  \end{tabular}
 \end{center}
$^\dagger$ --- Normalization condition $Q_n$=1.
\end{table}

\vspace{2cm}

\begin{table} [htb]
Table \ref{tabl4}. \\[1cm]
 \begin{center}
  \begin{tabular}{|rr|r@{.}l|r@{.}l|r@{.}l|r@{.}l|}
\hline \hline
\multicolumn{2}{|c}{ Normal } & \multicolumn{8}{|c|}{Normalized amplitudes} \\
    \cline{3-10}
\multicolumn{2}{|c}{ frequency/cm$^{-1}$}
   & \multicolumn{2}{|c}{\rule{0em}{1.5em} $(r_1-r_{s1})/$\AA}
   & \multicolumn{2}{|c}{$(\alpha-\alpha_s)/$rad.}
   & \multicolumn{2}{|c}{$(r_3-r_{s3})/$\AA}
   & \multicolumn{2}{|c|}{$(J_x-J_{sx})/J_{sz}$} \\
\hline
 $\omega_1$=& 2095.8 &   1 &$^\dagger$     &0 & 030 & 0 & 015 &    0 & 004 \\
 $\omega_2$=& 1271.9 &$-$0 & 022 &1 &$^\dagger$  & $-$0 & 009 &    0 & 093 \\
 $\omega_3$=& 2438.0 &$-$0 & 015 &0 & 003 & 1 &$^\dagger$     & $-$0 & 002 \\
 $\Omega_{xz}$=& 211.7 &   0 & 000 &0 & 429 & $-$0 & 009 &    1 &$^\dagger$  \\
\hline \hline
  \end{tabular}
 \end{center}
$^\dagger$ --- Normalization condition $Q_n$=1.
\label{pagemax}
\end{table}

\end{document}